# Genetic determinism of circadian rhythm of feed intake and relation with feed efficiency evaluated in group-housed growing Large White pigs.

Lucile Riaboff[1*], Ingrid David[1]

[1]GenPhySE, Université de Toulouse, INRAE, ENVT, 31326, Castanet-Tolosan, France

*Corresponding author

E-mail addresses:

LR: lucile.riaboff@inrae.fr

ID: ingrid.david@inrae.fr

# Abstract (maximum 350 words).

## Background

Genetic parameters of feeding behaviours traits from electronic feeding stations in relation to feed efficiency have been widely explored. However, genetic determinism of the circadian rhythm of feed intake throughout the fattening phase in group-housed growing pigs fed *ad libitum* has never been investigated, despite the well-known relationships between animals' circadian rhythms and the optimization of their metabolism. The objective of this study was to (i) propose three new traits derived from time-frequency approach applied to electronic feeding data from 2,297 Large White pigs that reflect the consistency of circadian feed intake rhythm throughout fattening (so called DayCR) and the precocity of its establishment (so called IndexCR and gCR), and then to (ii) estimate the heritability of those traits and their genetic correlations with residual feed intake using a multiple trait model.

## Results

Results highlighted moderate heritability estimates for the three circadian traits (range $h^2$: [0.24; 0.35]) and high heritability for residual feed intake (0.41). High genetic correlations (range of absolute values: [0.87; 0.98]) among circadian traits suggested that pigs exhibiting a 24-hour periodicity in feed intake on most days of fattening, particularly during the final fattening period, establish their circadian rhythm earlier than the other pigs. The low (range of absolute values: [0.18; 0.27]) but favourable genetic correlations between residual feed intake and circadian traits revealed that animals with a consistent and early 24-hour periodicity of feed intake also tend to be more feed efficient.

## Conclusions

This study proposed to apply time-frequency analysis on longitudinal feeding data to detect 24-hour periodicities in the hourly feed intake pattern across days throughout fattening in growing-pigs. Results suggested that part of the variability observed in the establishment of circadian rhythm is genetically driven, further supporting the feasibility of genetic selection on circadian traits. Considering the well-established biological mechanisms underlying circadian feeding rhythm, selecting animals for their ability to exhibit an early and consistent 24-hour periodicity of feed intake could promote metabolic homeostasis, thereby enhancing animal performance and resilience.

# Background

Development of technologies such as radio-frequency identification [1,2], camera vision system at the feed trough [3] or electronic feeding stations (EFS) [4] offered the opportunity to continuously measure feed intake in growing pigs. In that regard, feeding behaviours traits, such as daily feed intake, feeding rate, feeder occupation time, number of visits and visit duration, have been widely explored in relation to animal performance traits, health and welfare [5–9]. Genetic analysis showed moderate to high heritability estimates (ranging from 0.36 to 0.50) for most feeding behaviour traits [10,11]. Genetic correlations between feeding traits and performance or carcass traits were generally low, except with average daily gain [10,11] and residual feed intake (RFI) [12,13]. Recently, hourly feed intake derived from EFS has been used to explore the variation in the circadian rhythm of feed intake and diurnal patterns in growing-finishing pigs, defined as the recurrence of the feeding intake pattern along a 24-hour cycle [14]. Although the diurnal feeding pattern, *i.e.*, a small peak of feed intake at the beginning of the day and a larger peak at the end of the day, is commonly acknowledged as the dominant pattern in group-housed growing pigs fed *ad libitum* [15–17], substantial inter-individual variation in its expression was reported [8,14]. For instance, Gertheiss et al. [18] observed that some pigs exhibited a larger feeding peak in the morning than in the afternoon, or showed only a single main peak in the afternoon. In contrast, Ingram et al. [19] reported individuals feeding mainly during night-time hours. Additionally, Bus et al. [14] reported inter-individual variation in the repeatability of feeding behaviour over a 24-hour cycle, ranging from pigs exhibiting no detectable circadian rhythm of feed intake to those showing highly consistent feeding patterns from one day to another. This is consistent with the study by Maselyne et al [1] with pigs exhibiting a consistent number of visits and visit intervals throughout the entire growing period, while others showed high day-to-day variability. However, these findings do not align with the natural circadian rhythms in growing pigs which are expected to generate a consistent diurnal pattern of feed intake across days. Indeed, feeding behaviour in animals arises from an endogenous circadian clock entrained by environmental factors such as light and food, known as *zeitgebers*, that regulates anticipatory behavioural and physiological responses, such as promoting or inhibiting food intake. Inter individual variability in the circadian rhythm of feed intake may be of genetic origin. Indeed, Vahlhaus et al. [20] recently showed in human adult twins' cohorts that all the eating time components were highly or moderately heritable, with the circadian time of the first eating event exhibiting the highest heritability (0.59). Similarly, Lopez-Minguez et al. [21] reported genetic influences for food timing in adult twins' cohorts, especially for the timing of breakfast (heritability: 0.56) and the timing of lunch (heritability: 0.38). Furthermore, the timing of food intake also show strong and significant genetic correlations (0.78-0.91) with sleep timing and chronotype, suggesting that individual responses face to circadian cues are partly genetically-driven. Genetic determinism of the circadian rhythm of feed intake in group-housed growing pigs has yet to be investigated, despite its potential as a proxy of optimal metabolism. Indeed, fine-tuned synchronisation of the peripheral clocks over a 24-hour cycle contributes to efficient energy metabolism by promoting effective optimal nutrient assimilation, mobilization for various functions and discarding metabolic waste at specific times. In humans, consistent daily pattern of eating/fasting aligned with daily/light cycle actively contributes to maintain circadian physiology, while frequent disruptions in eating/fasting rhythms or genetic disruption of circadian clock predispose to obesity, type 2 diabetes, or cardiovascular diseases [22,23]. Observing regular daily pattern of food intake in animals is thereby one of the outward manifestations of optimal metabolism functioning. However, exploring the genetic basis of the circadian rhythm of feeding behaviour in growing pigs first requires the development of novel traits that can serve as proxies of the consistency and precocity of the 24-hour periodicity of feed intake across the fattening period.

The objective of this study is to (i) derive innovative traits reflecting the day-to-day consistency and precocity of the feed intake pattern in housed-growing Large White (LW) pigs fed *ad libitum* using data from EFS and then to (ii) estimate the genetic parameters of those traits. Furthermore, given the well-established relationship between consistent circadian feeding rhythms and optimal nutrient metabolism [23,24], we assume that pigs exhibiting early and consistent circadian rhythm of feed intake during the fattening period are also more feed-efficient; thereby the last objective of this paper is to (iii) investigate the phenotypic and genetic correlations between circadian traits and RFI.

# Methods

## Data collection

Data were collected in accordance with the applicable national regulations on livestock welfare in France.

Data used for this study were from 2,312 male LW pigs in the growing-finishing period of the INRAE France Génétique Porc facility [25] in Le Rheu, France, from 2017 to 2019. Piglets were born in selection farms from the breeding companies Axiom (Azay-sur-Indre, France) and Nucleus (Le Rheu, France) and arrived at approximately 62 ± 3 days of age (in the fattening unit at a body weight of 23.8 ± 4.4 kg). They were raised for approximately 100 days (range: [73; 123]) until they reached the target live weight of 115 kg. Pigs that left before 80 days were removed (n = 15), leading to 2,297 pigs for the analysis. For each batch, i.e., a group of pigs that arrived the same week at the station, couples of full sibs were separated and allotted in pens of 14 animals (range: [9; 15]). Over the entire period, animals were distributed into 52 batches and 440 pens (range of pens per batch: [6; 9]). Each pen was equipped with a single-place EFS with an integrated weighing scale (Genstar; Acemo Skiold, Pontivy, France) that recorded individual feed intake and body weight at each animal visit. Pigs used for this study were also involved in a trial where two sets of pigs were fed a two-phase dietary sequence. Therefore, 1571 animals (68%) were fed with a conventional (CO) dietary sequence formulated to cover energy and amino acids requirements of pigs while 726 animals (32%) received a less digestible diet with increased dietary fibre content (HF). The detailed composition of CO and HF feeds is described in [26]. Pigs had *ad libitum* access to feed and water at all stages of growth.

## Circadian traits and residual feed intake calculation

All data analyses were carried out with R, version 4.2.1 [27].

Entry time, exit time and individual feed intake as well as body weight were recorded for each visit to EFS. Visits were then merged when the interval between two consecutive visits was less than 15 seconds, which were considered as EFS errors. Feed intake per visit was then aggregated to get the feed intake for every hour, calculated as the sum of the feed intake from all visits within each hour (unit: kg/hour), for each pig over the entire fattening period [14]. Any hour with no recorded feed intake was assigned a value of 0 kg. We focused on feed intake rather than other traits measurable from EFS as its variation is relatively low, the results are easy to interpret, and it does not require a meal criterion to aggregate feeding visits into meals; thereby, no meal criterion was used in this study [7,14].

Wavelet analysis was applied on the hourly feed intake for each pig period to detect any circadian rhythms over the entire fattening period, as carried out in [14]. Wavelet analysis is a time-frequency approach commonly used to detect activity rhythmicity in mammals [28,29], including humans [30].

This method allows the detection of specific periodicities and repeating patterns by measuring the similarity (also called *wavelet power*) between the signal and a pre-chosen sinusoid (also called *base wavelet*) of different periods, translated across the time series. Detrending was first applied to the hourly feed intake for each individual to correct for the increase in average feed intake observed over the fattening period [14]. Indeed, such main trends may dominate the signal by carrying most the variance and power at low frequencies, thereby masking the periodicities of interest [31]. Detrending was applied by subtracting the local regression model (*loess* function; span = 0.75) fitted to the hourly intake for each animal. Furthermore, to avoid observing changes in wavelet power that are due solely to amplitude differences (e.g., older pigs consuming as much as younger pigs in far fewer visits) rather than true periodicities, correction for amplitude was also applied. For this purpose, each hourly feed intake data point was divided by the difference between the highest and lowest hourly feed intake of the 7 days-window that contains the data point [14]. Wavelet analysis was applied to the detrended and amplitude-corrected hourly feed intake over the entire fattening period using the *analyze.wavelet* function in the R package *WaveletComp* [32]. The default Morlet base wavelet was used. Time resolution was set to hourly data (*dt = 1/24*) and frequency resolution was fixed at 0.1 (*dj = 0.1*) to maintain high precision while avoiding excessive computations. The wavelet transform was restricted to periods between 12 and 48 hours (*lowerPeriod = 0.5*; *upperPeriod = 2*) to focus on the periodicity of interest (24 hours) and reduce computation time. For each pig and each hour, the wavelet power associated to the periodicity closest to 24 hours was extracted. Significance was assessed by comparing the observed wavelet power to that of 1,000 simulated time series (*n.sim = 1000*) generated from white noise (*method = "white.noise"*). The median p-value for each day and each pig was computed from the hourly p-values. A day was considered to have a significant 24-hour periodicity if its median p-value was below 0.05 [14]. The following three traits were calculated for 2,297 pigs to depict day-to-day consistency and precocity of the circadian feed intake establishment:

**DayCR**: The percentage of days where a significant 24-hour periodicity was detected. As a preliminary analysis from wavelet spectra and median-pvalue distribution revealed that the main differences between pigs occurred at the end of the fattening period, DayCR was actually calculated from day 67 (i.e., 67 days after pig's arrival) to day 99 (i.e., median departure day) of the growing-finishing period. DayCR trait depicted consistency of the 24-hour periodicity of feed intake in the final period of fattening.

**IndexCR**: IndexCR was defined as the first day on which a sliding 7-day window contained at least 5 days exhibiting a statistically significant circadian rhythm of feed intake. For each pig, the sliding window was applied on the entire fattening period regardless its duration. If no 24-hour periodicity was detectable on 5 of 7-days throughout the fattening period, then IndexCR value was set to the day of departure for that pig. It should be noted that several window sizes and thresholds were tested but the results were consistent across parameter settings (*data not shown*). IndexCR trait was indicative of the precocity of the establishment of the circadian feed intake.

**gCR:** Hierarchical Ascendant Classification (HAC) was applied to the percentage of days for which a 24-hour periodicity was detected in each of the following period of fattening: from day 1 to day 33, from day 34 to day 66, and from day 67 to day 99. Euclidean distance was used as the dissimilarity metric; *hclust* function was applied to get the clusters using ward agglomeration method based on the square of the Euclidean distance (*method = "ward.D2"*). Three main clusters emerged from the dendrogram obtained with the HAC (see Figure 2 (a)). Consolidation of the three clusters was then performed using k-means clustering, with the cluster averages from HAC used as initial centroids. The distribution of the percentage of days with a significant circadian rhythm of feed intake across the three periods of fattening revealed different precocity of circadian rhythm establishment (see Figure 2 (b)). Accordingly, the three clusters were renamed based on their precocity as "early", "intermediate" and "late"; thereby, gCR corresponded to the cluster to which the pig belonged.

Residual feed intake was calculated for each of the 2,297 LW pigs. It corresponds to the difference between observed Daily Feed Intake (DFI) and expected DFI for maintenance and production requirements [33]. It was derived from the residuals of a multiple linear regression of average DFI on average daily gain (requirements for growth), lean meat percentage, backfat thickness and average metabolic body weight (AMBW) computed as:

$$AMBW = \frac{(\mathrm{BWe}^{1.6} - \mathrm{BWs}^{1.6})}{1.6 \text{ x } (\mathrm{BWe} - \mathrm{BWs})}$$

Where BWs and BWe are the body weight at the start and end of the test period, respectively.

## Genetic analyses

The traits considered in the genetic analysis were DayCR, IndexCR, gCR and RFI, as defined in the previous section. gCR was treated as a continuous variable to reflect the gradual decrease in the percentage of days with a significant circadian rhythm from the cluster "early" to the cluster "late", (*i.e.*, "early" = 1 < "intermediate" = 2 < 'late" = 3). To facilitate model convergence, only RFI records of animals fed a conventional regime were considered. A multi-trait linear mixed model was used to jointly analyse the four traits of interest, allowing estimation of trait-specific variances and co-variances between traits. The model for trait *j* (RFI, DayCR, IndexCR, gCR) is:

$$\mathbf{y}_j = \mathbf{X}_j\mathbf{b}_j + \mathbf{Z}_j\mathbf{u}_j + \mathbf{W}_j\mathbf{c}_j + \mathbf{e}_j$$

where $\mathbf{y}_j$ is the vector of observation for trait $j$, $\mathbf{b}_j$ is the vector of fixed effects for the considered trait, $\mathbf{u}_j$ is the vector of random additive genetic values for trait $j$, $\mathbf{c}_j$ is the vector of random pen within batch contemporary group effect, $\mathbf{e}_j$ is the residual random effect, $\mathbf{X}_j, \mathbf{Z}_j$ and $\mathbf{W}_j$ are known incidence matrices. All random effects were independent from each other and distributed as centered normal distributions with variance covariance matrices equal to:

$$\begin{bmatrix} \sigma^2_{u,RFI} & & & symm \\ \sigma_{u,RFI-DayCR} & \sigma^2_{u,DayCR} & & \\ \sigma_{u,RFI-IndexCR} & \sigma_{u,DayCR-IndexCR} & \sigma^2_{u,IndexCR} & \\ \sigma_{u,RFI-gCR} & \sigma_{u,DayCR-gCR} & \sigma_{u,IndexCR-gCR} & \sigma^2_{u,gCR} \end{bmatrix} \otimes \mathbf{A}$$

for the genetic effects where **A** is the pedigree relationship matrix,

$$\begin{bmatrix} \sigma^2_{c,RFI} & & & symm \\ \sigma_{c,RFI-DayCR} & \sigma^2_{c,DayCR} & & \\ \sigma_{c,RFI-IndexCR} & \sigma_{c,DayCR-IndexCR} & \sigma^2_{c,IndexCR} & \\ \sigma_{c,RFI-gCR} & \sigma_{c,DayCR-gCR} & \sigma_{c,IndexCR-gCR} & \sigma^2_{c,gCR} \end{bmatrix} \otimes \mathbf{I}$$

and

$$\begin{bmatrix} \sigma^2_{e,RFI} & & & symm \\ \sigma_{e,RFI-DayCR} & \sigma^2_{e,DayCR} & & \\ \sigma_{e,RFI-IndexCR} & \sigma_{e,DayCR-IndexCR} & \sigma^2_{e,IndexCR} & \\ \sigma_{e,RFI-gCR} & \sigma_{e,DayCR-gCR} & \sigma_{e,IndexCR-gCR} & \sigma^2_{e,gCR} \end{bmatrix} \otimes \mathbf{I}$$

for the contemporary group effect and residuals, respectively, where **I** correspond to identity matrix of appropriate size.

Fixed effects included in the model were the age of the piglets upon arrival at the station and the number of days in the last period of fattening to account for shorter periods due to early departures, both included as a co-variable and the diet (CO or HF) for circadian traits. Variance components were estimated by the restricted maximum likelihood (ReML) method using the ASREML 4.2 software [34]. Heritabilities were calculated as the ratio of the genetic ($\sigma^2_u$) to the total variance ($\sigma^2_T = \sigma^2_u + \sigma^2_c + \sigma^2_e$)

# Results

## Inter-individual variability of the circadian feed intake traits

Distributions of DayCR and indexCR traits are displayed in Figure 1. Zero-inflated uniform distribution of DayCR suggested that most pigs did not exhibit a consistent circadian rhythm of feed intake, even during the last period of fattening (days 67 to 99). However, substantial inter-individual variability was observed for this trait (26.61% ± 26.79), ranging from 0% to 100% of days with a significant 24-hour periodicity detected. The distribution of IndexCR showed a peak around day 100, corresponding to the average default value assigned to pigs that never exhibited 5 out of 7 days with a significant circadian rhythm. IndexCR also showed substantial variation between individuals (71.63 ± 33.14), with some pigs establishing their first steady period as early as 7 days after arrival, while others never did throughout fattening. The secondary peak observed around 10 days confirmed the existence of distinct individual profiles in the precocity of the circadian feed intake establishment.

**Figure 1. Distribution (n = 2,297) of DayCR (a) and IndexCR (b) traits**

Legend: DayCR represents the percentage of days with a significant circadian rhythm of feed intake ($p < 0.05$) in the final fattening period. IndexCR is defined as the first day on which a significant circadian rhythm ($p < 0.05$) was observed for at least 5 of 7 consecutive days.

Three clusters emerged from the dendrogram resulting from the HAC (see Figure 2 (a)), each characterized by distinct distributions of the percentage of days with a significant circadian rhythm across the first (days 1-33), second (days 34-66), and last (days 67-99) periods of the growing-finishing (see Figure 2 (b)). The main cluster (h = 732) consisted of 1,496 pigs (65%) showing almost no circadian rhythmicity, regardless of the period. Indeed, 24-hour periodicity was detected for only 6.2% ± 10.2% of the days in the first period, 6.9% ± 10.9% in the second, and 10.3% ± 10.0% in the last one. Individual wavelet spectra for this cluster confirmed the absence of periodicity near the 24-hour period (Figure 3 (a); pig 657), although some animals sporadically exhibited periods with a circadian rhythm (Figure 3 (a); pig 780, 997). This cluster was called late thereafter to denote a late (or absent) establishment of circadian rhythm of feed intake. The second cluster (h = 616) consisted of 545 pigs (24%) showing a gradual increase of the number of days with a significant circadian rhythm between the first (9.4% ± 11.4%), second (19.5% ± 15.2%) and last period of fattening (54.5% ± 18.0%). However, individual wavelet spectra in this cluster revealed inter-individual variability. While some pigs exhibited a gradual increase in the number of days with 24-hour periodicity (Figure 3 (b); pig 380), others displayed a more sporadic rhythm, with intervals where a circadian rhythm was clearly visible, interspersed with periods where the 24-hour periodicity was transiently lost (Figure 3 (b); pig 924, 166). This cluster was called intermediate later on. The third cluster (h = 391) included 256 animals (11%) that exhibited a progressive increase in the percentage of days with a 24-hour period detected, from 36.8% ± 23.2% of days in the first period to 64.6% ± 18.3% in the second period, before reaching a plateau at 62.5% ± 25.8% in the last one. However, although individual wavelet spectra confirmed a more consistent circadian rhythm of feed intake throughout fattening (Figure 3 (c); pig 648), some animals from this cluster also transiently lost the 24-hour periodicity of feed intake during fattening (Figure 3 (c); pig 079, 098). This cluster was called early to depict early establishment of circadian rhythm.

**Figure 2. (a) Dendrogram obtained from the Hierarchical Ascendant Clustering applied on the percentage of days with a significant circadian rhythm ($p < 0.05$) in the first [day 1 to 33], second [day 34 to 66] and last [day 67 to 99] period of fattening. (b) Distribution of the percentage of days with a significant circadian rhythm of feed intake ($p < 0.05$) in the late, early and intermediate clusters (from left to right) during the first [day 1 to 33], second [day 34 to 66], and last [day 67 to 99] period of fattening.**

Legend: The gCR trait corresponds to the three main clusters, marked in light brown, green, and dark brown, which are respectively defined as late, early, and intermediate.

Finally, the hourly average feed intake across pigs during the final fattening period (days 67-99) revealed distinct patterns over a 24 hour-period depending on the cluster. Pigs in the late cluster (see Figure 4 (a)) exhibited a relatively continuous feeding pattern throughout the day, whereas the characteristic two peaks of the diurnal pattern were starting to emerge in the intermediate cluster (see Figure 4 (b)). In the early cluster, the two peaks of feed intake in the morning and late afternoon appeared clearly (see Figure 4 (c)).

**Figure 3. Wavelet spectra over the entire growing-finishing period for three example pigs from each cluster, as defined in the gCR trait: (a) late, (b) intermediate, and (c) early.**

Legend: Pig numbers are shown in the grey box above each spectrum. The black line indicates the most likely periodicity, the white lines delimit the periodicity significantly stronger than white noise ($p < 0.05$), and the colours represent the strength of the periodicity (wavelet power level).

**Figure 4. Hourly feed intake (g) averaged across days and pigs during the last period of the growing-finishing phase [day 67 to 99], for the (a) late, (b) intermediate and (c) early clusters.**

## Estimation of the genetic parameters of the circadian traits and residual feed intake

Heritability, genetic and phenotypic correlations between the different traits are reported in Table 1. Estimated heritabilities of the three circadian traits were moderate, ranging from 0.24 ± 0.05 for IndexCR to 0.35 ± 0.06 for DayCR. Estimated heritability of RFI was high (0.41 ± 0.07). Genetic correlations were high and agonist between all circadian traits. Genetic correlation between gCR was equal to 0.93 ± 0.04 with IndexCR and -0.98 ± 0.02 with DayCR. Genetic correlation between DayCR and IndexCR was equal to -0.87 ± 0.06. Genetic correlation estimates between the three circadian traits and RFI were low to moderate, ranging from 0.18 to 0.27 in absolute value, all tending in the same direction. Phenotypic correlations aligned in direction with the genetic correlations, though they were of lower magnitude, ranging from 0.52 to 0.80 (in absolute value) between circadian traits and from 0.005 to 0.07 (absolute value) between RFI and circadian traits.

**Table 1. Estimates of heritabilities, genetic and phenotypic correlations for the circadian feed intake traits (DayCR, IndexCR, gCR) and residual feed intake (RFI).**

Legend: On diagonal (bold): Heritabilities ± SE; above the diagonal: genetic correlations ± SE; below the diagonal: phenotypic correlations ± SE

# Discussion

## Innovative approach to explore periodicities of feeding behaviour from EFS data

Non-circadian feeding behaviour traits from EFS data in pigs were actually widely investigated in the literature [5,10,11,13]. However, the genetic determinism of 24-hour periodicity of the hourly feed intake has never been explored. In our study, we proposed three innovative traits to depict the consistency and precocity of the circadian feed intake pattern during fattening. These traits were derived from the wavelet analysis applied on the hourly feed intake for each pig across the entire growing-finishing phase. Unlike the Short-Time Fourier Transform, wavelet analysis uses variable window sizes, providing both high time resolution for short periods and high frequency resolution for long periods. In this way, this method allows the capture of multiple periodicities and their temporal variations, making it especially suitable for signals with complex or non-stationary patterns [35]. This method, applied for the first time on EFS data in pigs in Bus et al. [14], has previously been used to detect activity rhythmicity in mammals [28,29], including humans [30]. This method could also be applied to investigate 24-hour rhythm in other feeding behaviour traits (e.g., feeding visits, feeding rate, occupation time) or to detect other feed intake rhythms, such as the 12-hour periodicity associated with the day/night cycle. This approach could also be deployed to investigate periodicities in any longitudinal data, such as activity in growing pigs [36] or postural behaviours in lactating sows [37]. Therefore, the methodology promoted in our study opens avenues for the identification of new traits based on periodicities in longitudinal data and for assessing their genetic basis.

## Genetic determinism of circadian rhythm of feed intake and correlation with feed efficiency

The proportion of days exhibiting a significant circadian rhythm in feed intake increased progressively over the growing-finishing period. This result was also observed in Bus et al. [14] and is coherent with the development of the circadian pattern with age in mammals [38]. In pigs, the diurnal pattern of feed intake is weak in young animals and becomes more distinct with age [15]. However, high inter-individual variability was found for the three circadian traits. Indeed, the percentage of days with a significant 24-hour periodicity of feed intake in the last period of fattening ranged from 0% to 100%, as also observed in Bus et al. [14]. The onset of a steady circadian rhythm during the growing-finishing phase was also highly variable, ranging from as early as 7 days after arrival to not occurring at all. Unsupervised clustering also revealed three main profiles. A first profile comprised animals that showed almost no 24-hour periodicity throughout the entire growing-finishing period (called *late*) while a second one included animals that displayed a clear circadian pattern in approximately 1/3 of the days at the start of the fattening and up to 2/3 of the days from the second period (called *early*). Between these extremes, a third profile exhibited a more moderate increase in circadian rhythmicity across the fattening phase, interspersed with several non-rhythmic episodes (called *intermediate*). This finding is coherent with the study of Bus et al. [14] in which the wavelet spectra analysis revealed that pigs ranged from exhibiting a steady circadian rhythm throughout the growing-finishing phase to showing no distinguishable significant rhythm, with intermediate individuals alternating between rhythmic and non-rhythmic periods. Interestingly, a preliminary analysis of the distribution of feed intake across the day in the final fattening period for the three main profiles highlighted a more pronounced diurnal pattern in early pigs. This suggests that the 24-hour periodicity observed at the end of fattening results from the consistent repetition of the diurnal feeding pattern across days. In contrast, the late pigs ate more evenly throughout the day without any clear pattern of feed intake, which may explain the absence of any 24-hour periodicity. This is also consistent with the correlations found in Bus et al. [14] between the proportion of days with a circadian rhythm, the lowest intake, the proportion of intake obtained at night and the highest probability to start eating, indicating that pigs with well-defined circadian rhythms maintain similar patterns across days and avoid feeding at night.

The moderate heritability of circadian traits highlighted that part of the variability observed between individuals is explained by genetics. These values actually fall within the range of heritabilities reported for the common feeding behaviour traits, such as the number of visit/day, the daily feeding time, the daily feed intake, the average feed intake per visit, etc [11,13,36]. The genetic correlations between the three circadian traits explored in our study were strong as well as the corresponding phenotypic correlations, suggesting that gCR, IndexCR and DayCR are actually the same trait. Finally, we found a favourable genetic correlation between circadian traits and RFI, as also observed in the literature with other feeding traits [13,39]. This result is actually coherent with our initial hypothesis that pigs exhibiting early and consistent circadian rhythm of feed intake during the fattening period are also more feed-efficient.

## Limitations of the study

Our study focused on the presence or absence of a significant 24-hour periodicity detected in the hourly feed intake of group-housed growing pigs during fattening. However, the 24-hour feed intake pattern itself has not been investigated, although it may be even more relevant. In fact, a 24-hour periodicity of feed intake was detected in only 20% of the days across the entire fattening, with 65% belonging to the late cluster. This indicates that most pigs did not exhibit a repeatable feed intake pattern during fattening. In that regard, Bus et al. [14] modelled diurnal feed intake patterns from 98 pigs and reported a range of patterns that changed with age. A hypothesis is that growing pigs may be too young to establish a consistent feed intake pattern from one day to another, thus preventing investigation of the genetic basis of the pattern itself. Another limitation is that we did not take into account in the data filtering the potential perturbations (environmental, health events, etc.) experienced during the growing-finishing phase because of the low reliability of such information (e.g., national outdoor climatic data are not representative of the indoor environment, health data at the batch or pen level do not accurately reflect the sanitary status of individual animal [40]). As it is well known that perturbations can alter circadian patterns in livestock [41], the delayed establishment of a circadian rhythm of feed intake or the non-rhythmic periods interspersed between stable phases may result from uncontrolled environmental factors, thereby biasing genetic analyses. However, it is worth noting that similar findings were also observed in Bus et al. [14] whereas days surrounding diseases or social stressful events were removed from the study. The stability and precocity of the circadian rhythm of feed intake could also be partly influenced by the social hierarchy among pigs [8,14]. Indeed, pig feeding behaviour at daily and diurnal levels is affected by competition at the feeder [42,43]; with 14 pigs per feeding space, the threshold for compromised daily feed intake during peak hours may be reached, as changes in feeding behaviour were observed from 15 pigs per feeding space [44]. It was already reported that dominant pigs have easier access to the feeder, particularly during peak hours [45]; therefore, subordinate pigs may be forced to spread their feed intake throughout the day. Consequently, a 24-hour periodicity is more likely to be detected in high-ranking pigs than in low-ranking ones. However, it is worth noting that in our study, transient losses of 24-hour periodicity were also observed in pigs that showed a circadian rhythm early in the fattening period. Since no pen reallocation occurred during growing-finishing, these non-rhythmic periods are unlikely to be caused by social hierarchy. Additionally, although it has been reported that low-ranking pigs visit the EFS more frequently but for shorter durations [45], this does not necessarily disrupt a 24-hour periodicity, as long as their feeding pattern remains consistent from day to day.

## Perspective for genetic selection on circadian feed intake traits

Estimation of the variance components for the three circadian traits suggested that genetic selection could be considered. Beyond the favourable correlation found with RFI, selecting pigs for their ability to exhibit circadian rhythm of feed intake early in the growing-finishing phase could actually lead to underlying benefits. Indeed, the detection of a 24-hour periodicity in feed intake may indicate that glucocorticoid and melatonin secretion are well aligned with external cues, particularly the light-dark cycle, and that core clock proteins, including CLOCK and BMAL1, are properly regulated through the feedback loop involving PER and CRY proteins. This fine-tuned regulation allows coordination of anticipatory behavioural and physiological responses, including the timing of food intake across the day [46]. For example, Boumans et al. [47] showed through simulations validated with empirical data that any shift in the peak of cortisol or melatonin alters the pattern of feed intake. Bechtold et al. [48] also reported that global disruption of clock genes leads to a loss or alteration of feeding rhythms. In turns, feeding schedules also exert a strong influence on circadian clocks in peripheral organs [46]. Disruption of the 24-hour feeding rhythm further leads to metabolic consequence. For example, Van Erp et al. [49] induced circadian misalignment of feed intake in pigs through night feeding and observed increase fat deposition and decrease net carbohydrate oxidation compared to the group fed *ad libitum* during daily hours. In addition, Gilbert et al. [50] reported significant changes in allele frequencies for two genes involved in circadian rhythm regulation (ARNTL and CLOCK) in the pig line genetically selected for lower feed efficiency. These recent findings suggest that genetic variability in feed efficiency in pigs may be linked to alterations in circadian rhythm. Supported by the favourable correlation between circadian traits and RFI observed in our study, it is reasonable to suggest that the 24-hour periodicity of feed intake, arising from the repeated diurnal pattern of feed intake, could serve as a proxy for optimal nutrient metabolism. Therefore, circadian traits could be used instead of RFI, for instance in lactating sows where its calculation may be challenging [51]. In that case, data source such as video analysis to derive sow activity [37] could be used to estimate circadian traits. Moreover, insights into resilience indicate that animals experiencing positive affective states tend to maintain stable feeding patterns [8,52]. This corroborates findings obtained in Gilbert et al. [12] where pigs' responses to stress or challenges were tested in divergent genetic lines selected for low/high feed efficiency. Indeed, although the resource allocation theory suggests that pigs genetically selected for higher feed efficiency may have a reduced capacity to cope with stressors [53], due to an altered ability to reallocate nutrients for stress and defence responses under environmental challenges, Gilbert et al. [12] concluded that none of the experiments support this theory. In contrast, between 10 weeks of age and slaughter, 1.8-fold fewer pigs were culled in the high feed efficiency line in the first seven generations of selection [12]. Similarly, significant lower high scores of lameness, leg lesions, bursitis and tail lesions were observed in the genetic line selected for high feed efficiency [36]. During heat stress, Campos et al. [54] also reported that highly feed-efficient pigs initiated acclimation responses more rapidly than less efficient ones. We can therefore assume that animals able to maintain a 24-hour periodicity of feed intake across most days of fattening are likely more resilient to stressful events. Therefore, our study opens new perspectives for genetic selection on circadian traits as potential proxies of metabolic homeostasis, considering the favourable outcomes expected for production performance and animal resilience.

# Conclusions

This study aimed to evaluate the genetic determinism of the circadian rhythm of the hourly feed intake in group-housed growing-finishing Large-White pigs from new circadian traits calculated from EFS data. The results encourage investigation into the genetic determinism of the circadian rhythm in young animals as well as in adults, as a promising new trait to integrate in pig selection. Indeed, empirical studies on the relationship between feed efficiency and resilience, together with the well-established biological mechanisms underlying circadian feeding rhythms, suggest that selection for consistent feeding circadian rhythms could improve metabolic homeostasis in animals, ultimately enhancing performance and resilience.

### List of abbreviations

AMBW: Average Metabolic Body Weight

CO: Conventional diet

EFS: electronic feeding stations

HAC: Hierarchical Ascendant Classification

HF: High-fibre diet

LW: Large White

RFI: residual feed intake

SE: Standard-Error

# Declarations

(please see our editorial policies for more information: https://www.biomedcentral.com/getpublished/editorial-policies#ethics+and+consent). If any of the sections are not relevant to your manuscript, please include the heading and write 'Not applicable' for that section.

### Ethics approval and consent to participate

Not applicable

### Consent for publication

Not applicable

### Availability of data and materials

The data that support the findings of this study were provided by Alliance R&D and are available from the authors upon reasonable request. Data requests should be addressed to the corresponding author.

# Competing interests

The authors declare that they have no competing interests

**Funding**

No funding was received

# Authors' contributions

LR: Methodology, Analyses, Writing - Original Draft

ID: Data Curation, Methodology, Analyses, Writing - Review & Editing, Project administration, Supervision

All authors have read and approved the final manuscript.

# Acknowledgements

The authors thank Alliance R&D (association composed of Axiom and Nucleus breeding companies and of IFIP, the French institute for pig and pork industry) and UE3P experimental unit for animal raising, data collection and sharing for this study. We also thank Suzanne Harari (PhD student; university of Paris Saclay, INRAE & university of Caen) for sharing R codes and expertise to facilitate downstream analysis of ASReml outputs.

**Author's information**

(optional)

# Figures

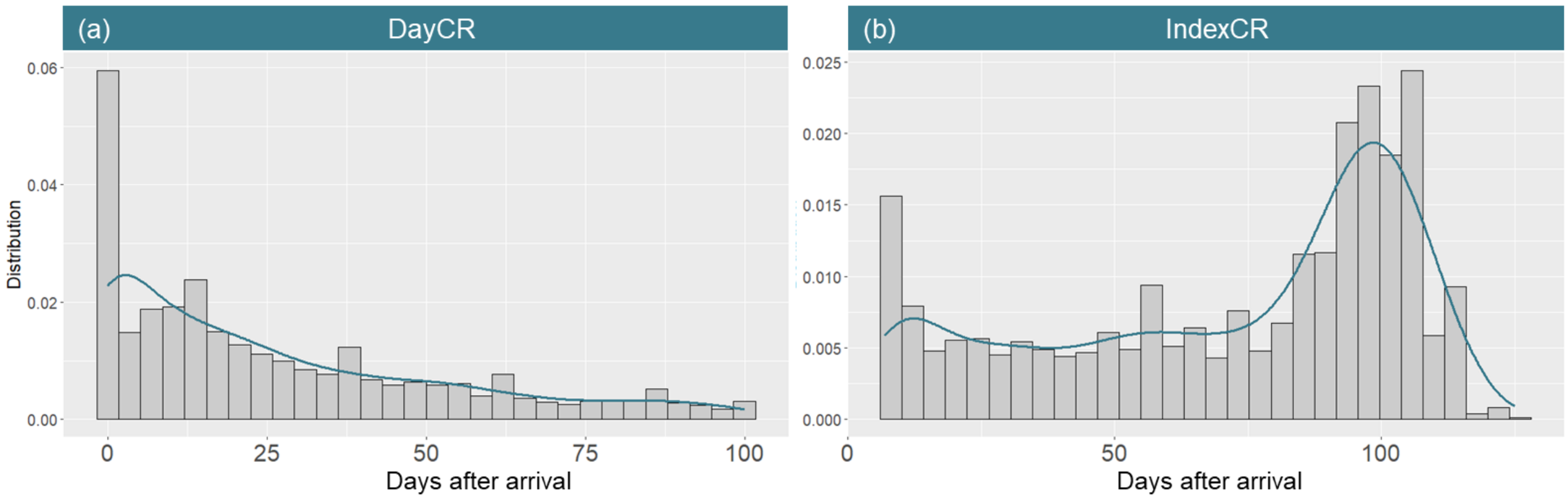

**Figure 1. Distribution (n = 2,297) of DayCR (a) and IndexCR (b) traits**

Legend: DayCR represents the percentage of days with a significant circadian rhythm of feed intake ($p < 0.05$) in the final fattening period. IndexCR is defined as the first day on which a significant circadian rhythm ($p < 0.05$) was observed for at least 5 of 7 consecutive days.

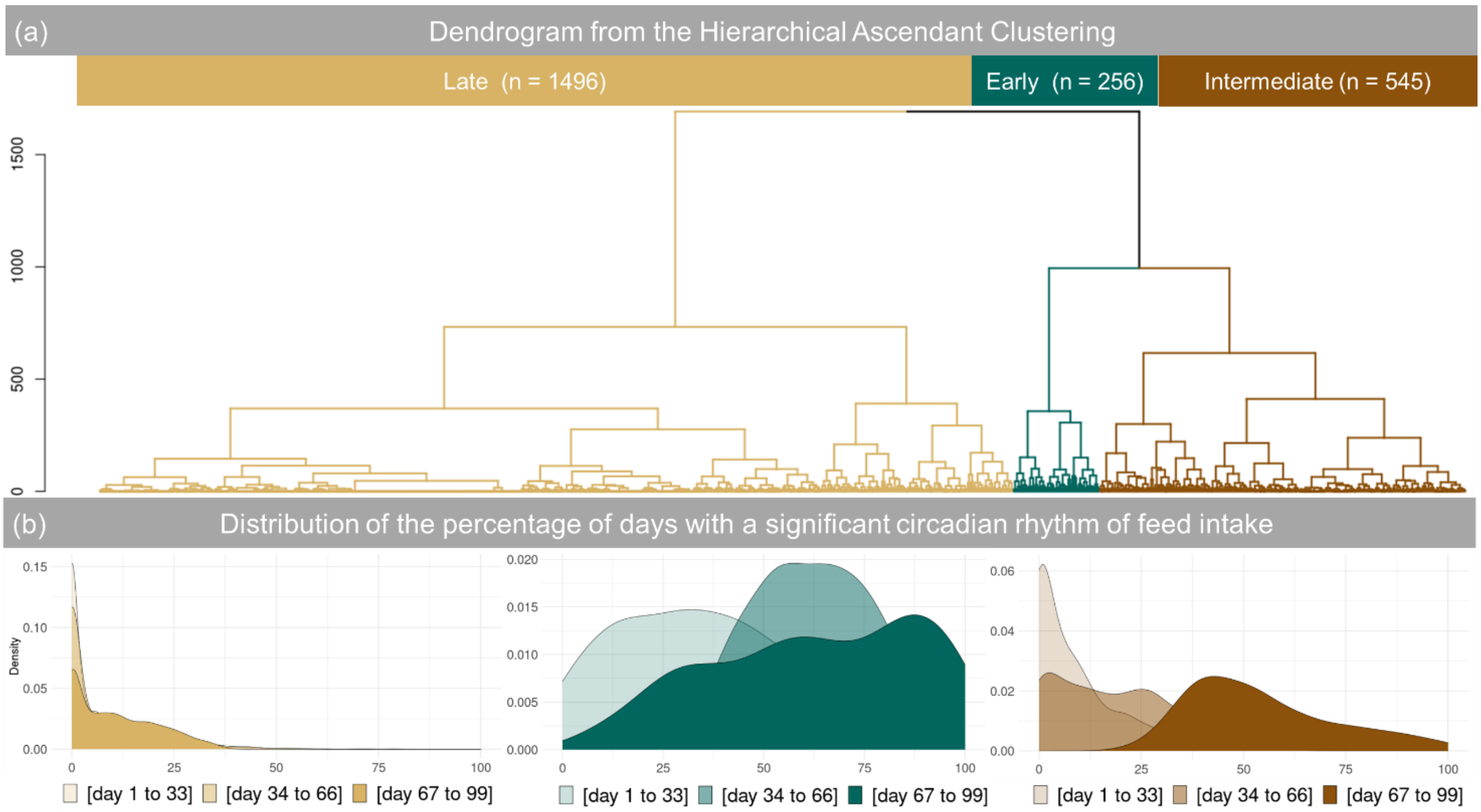

**Figure 2. (a) Dendrogram obtained from the Hierarchical Ascendant Clustering applied on the percentage of days with a significant circadian rhythm ($p < 0.05$) in the first [day 1 to 33], second [day 34 to 66] and last [day 67 to 99] period of fattening. (b) Distribution of the percentage of days with a significant circadian rhythm of feed intake ($p < 0.05$) in the late, early and intermediate clusters (from left to right) during the first [day 1 to 33], second [day 34 to 66], and last [day 67 to 99] period of fattening.**

Legend: The gCR trait corresponds to the three main clusters, marked in light brown, green, and dark brown, which are respectively defined as late, early, and intermediate.

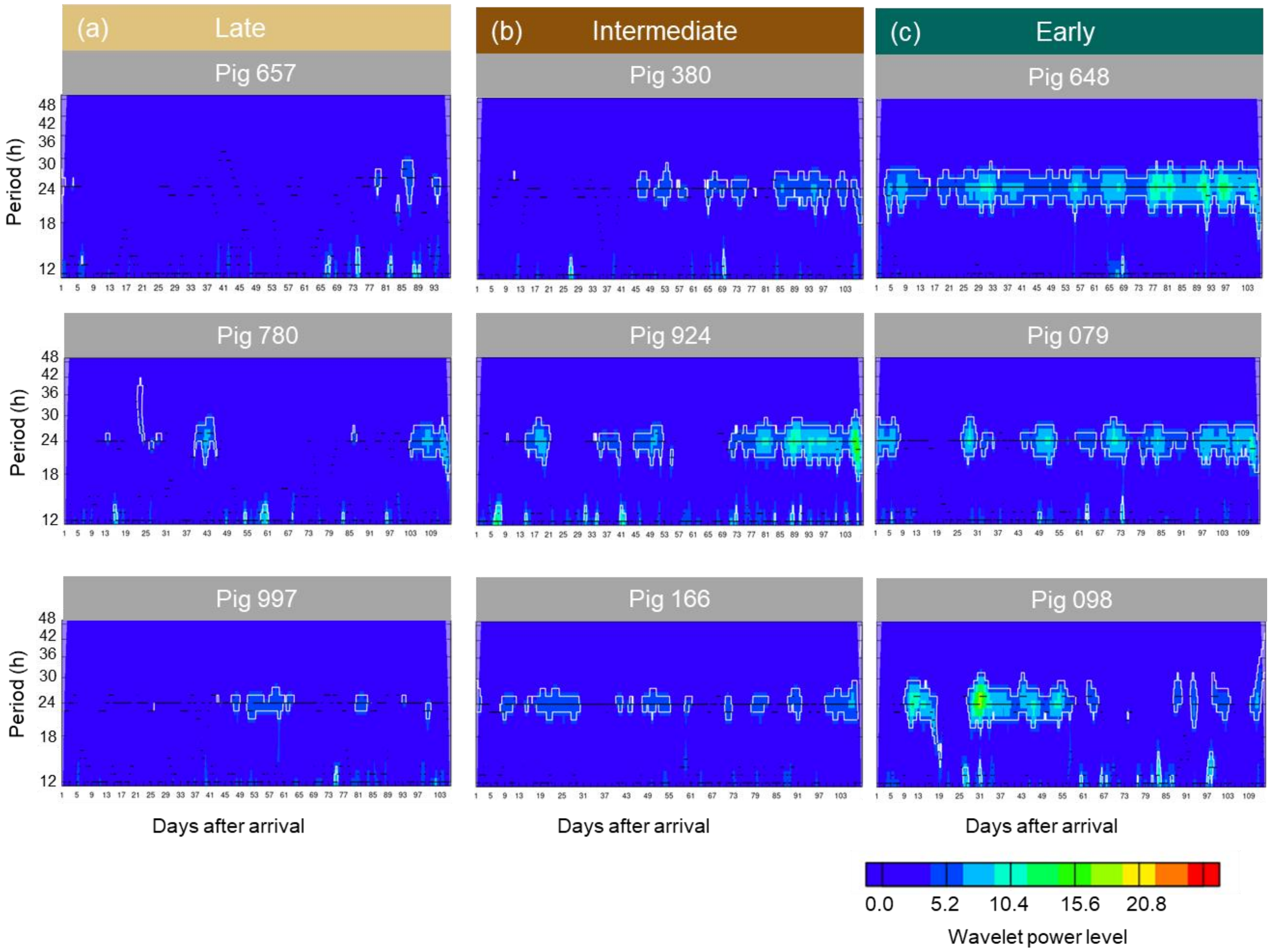

**Figure 3. Wavelet spectra over the entire growing-finishing period for three example pigs from each cluster, as defined in the gCR trait: (a) late, (b) intermediate, and (c) early.**

Legend: Pig numbers are shown in the grey box above each spectrum. The black line indicates the most likely periodicity, the white lines delimit the periodicity significantly stronger than white noise ($p < 0.05$), and the colours represent the strength of the periodicity (wavelet power level).

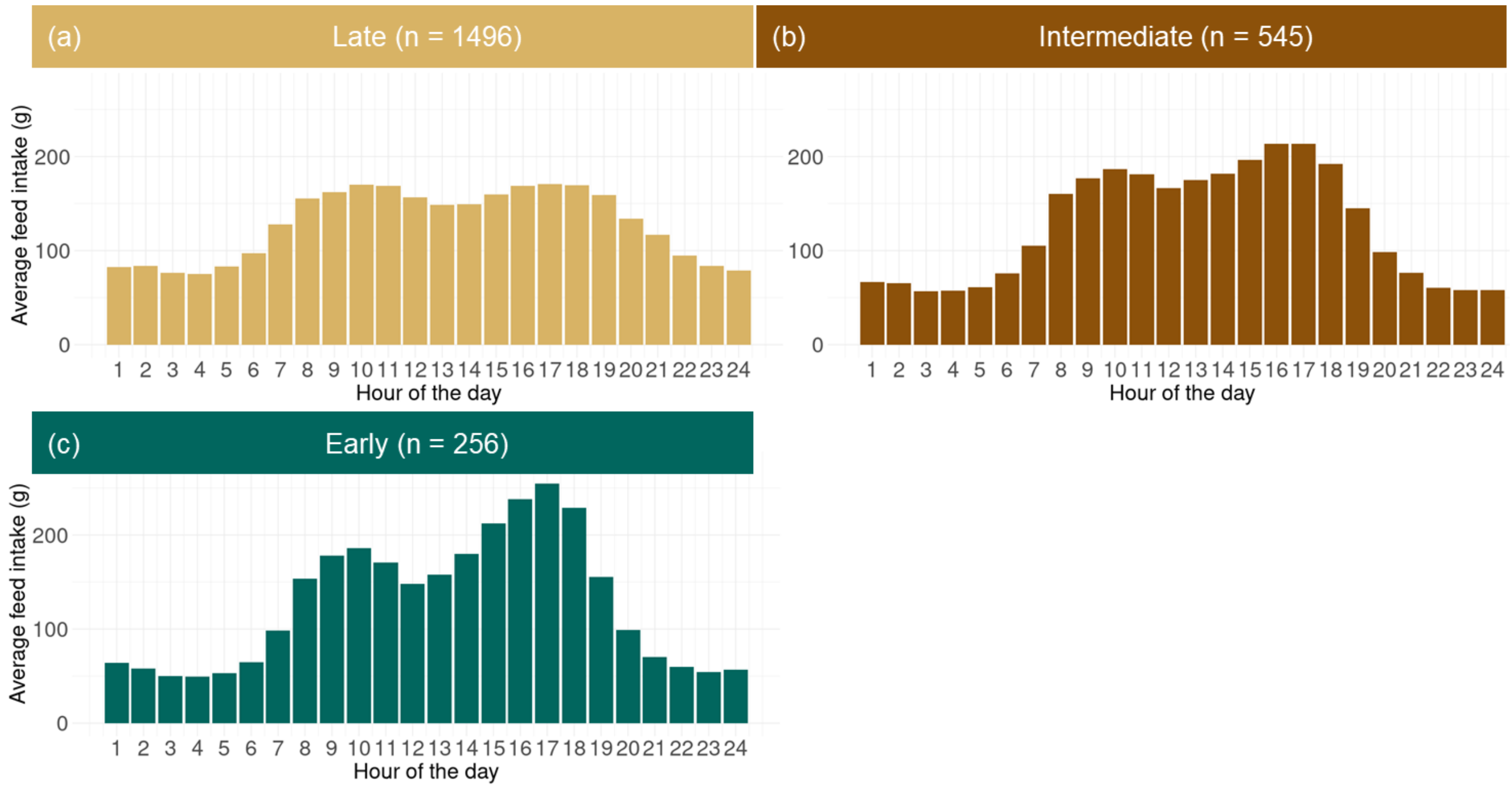

**Figure 4. Hourly feed intake (g) averaged across days and pigs during the last period of the growing-finishing phase [day 67 to 99], for the (a) late, (b) intermediate and (c) early clusters.**

# Tables

**Table 1. Estimates of heritabilities, genetic and phenotypic correlations for the circadian feed intake traits (DayCR, IndexCR, gCR) and residual feed intake (RFI).**

| | DayCR | IndexCR | gCR | RFI |
|---|---|---|---|---|
| DayCR | **0.35 ± 0.06** | -0.87 ± 0.06 | -0.98 ± 0.02 | -0.18 ± 0.12 |
| IndexCR | -0.52 ± 0.01 | **0.24 ± 0.05** | 0.93 ± 0.04 | 0.23 ± 0.14 |
| gCR | -0.80 ± 0.01 | 0.61 ± 0.01 | **0.29 ± 0.05** | 0.27 ± 0.13 |
| RFI | -0.005 ± 0.03 | 0.07 ± 0.03 | 0.01 ± 0.03 | **0.41 ± 0.07** |

Legend: On diagonal (bold): Heritabilities ± SE; above the diagonal: genetic correlations ± SE; below the diagonal: phenotypic correlations ± SE